\documentclass[prd,twocolumn,amsmath,amssymb,showpacs,floatfix,nofootinbib,preprintnumbers,superscriptaddress]{revtex4-1}
\usepackage[latin1]{inputenc}
\usepackage{amsmath}
\usepackage{amssymb}
\usepackage{graphicx}
\usepackage{dcolumn}
\usepackage{subcaption}
\usepackage{bm}
\usepackage{latexsym}
\usepackage{epsfig}
\usepackage[normalem]{ulem}
\usepackage{graphicx}
\usepackage[colorlinks=true]{hyperref}
\usepackage[dvipsnames]{xcolor}

\begin{document}
\title{How to measure CMB spectral distortions with an  imaging telescope}
\author{Suvodip Mukherjee}\email{mukherje@iap.fr}
\affiliation{Institut d'Astrophysique de Paris, UMR 7095, CNRS, 98 bis boulevard Arago, F-75014 Paris, France}
\affiliation{ Institut Lagrange de Paris, Sorbonne Universit\'es, 98 bis boulevard Arago, F-75014 Paris, France}
\author{Joseph Silk}\email{silk@iap.fr}
\affiliation{Institut d'Astrophysique de Paris, UMR 7095, CNRS, 98 bis boulevard Arago, F-75014 Paris, France}
\affiliation{ Institut Lagrange de Paris, Sorbonne Universit\'es, 98 bis boulevard Arago, F-75014 Paris, France}
\affiliation{The Johns Hopkins University, Department of Physics \& Astronomy, 3400 N. Charles Street, Baltimore, MD 21218, USA}
\affiliation{Beecroft Institute for Cosmology and Particle Astrophysics, University of Oxford, Keble Road, Oxford OX1 3RH, UK}
\author{Benjamin D. Wandelt}\email{bwandelt@iap.fr}
\affiliation{Institut d'Astrophysique de Paris, UMR 7095, CNRS, 98 bis boulevard Arago, F-75014 Paris, France}
\affiliation{ Institut Lagrange de Paris, Sorbonne Universit\'es, 98 bis boulevard Arago, F-75014 Paris, France}
\affiliation{Center for Computational Astrophysics, Flatiron Institute, 162 5th Avenue, 10010, New York, NY, USA}
\date{\today}
\begin{abstract}
We propose the implementation of an imaging telescope in combination  with an  inter-frequency calibrator to measure the spectral shape of the microwave sky by exploiting the differences in the sky intensity between multiple pairs of frequency channels. By jointly sampling the cosmological and foreground parameters in a Bayesian framework for $600$ instrument configurations, we determine the minimum calibration accuracy required in order to obtain measurements of spectral distortions and simultaneously measure spectral and spatial fluctuations of the CMB. We demonstrate the feasibility of this technique for a CMB mission like PICO (Probe of Inflation and Cosmic Origins), and show that a $10$-$\sigma$ measurement of the $y$-distortion along with  a two orders of magnitude improvement  on the FIRAS (Far-Infrared Absolute Spectrophotometer) $\mu$-distortion limit is feasible from this technique. We argue that longer term applications may be envisaged at even higher sensitivity, capable of attaining the $\mu$-distortion that provide a robust prediction of the $\Lambda$CDM model and even primordial recombination lines of hydrogen and helium,  in the context of the ESA (European Space Agency) Voyage 2035-2050 program. 
\end{abstract}
\maketitle

\section{Introduction} 
 {The first accurate measurements of the microwave sky spectrum established its blackbody spectrum with a thermodynamic temperature $2.736 \pm 0.017$ K. \cite{PhysRevLett.65.537}, soon followed by the} hitherto 
 unsurpassed Far-Infrared Absolute Spectrophotometer (FIRAS) \cite{1994ApJ...420..439M, Fixsen:1996nj}, and  later re-calibrated with the Wilkinson Microwave Anisotropy Probe (WMAP)  \cite{2009ApJ...707..916F} to obtain  the current best-fit value of the CMB thermodynamic temperature of $2.72548 \pm 0.00057$ \cite{2009ApJ...707..916F}, with no significant departure from a blackbody spectrum.

However, the standard Lambda Cold Dark Matter (LCDM) model of cosmology predicts guaranteed distortions   \cite{1970Ap&SS...7....3S,1994ApJ...430L...5H,Chluba:2016bvg, Chluba:2019kpb,2019arXiv190901593C} in the blackbody spectrum of the CMB due to several astrophysical and cosmological phenomena over a vast range of redshifts. The dominant source of late time ($z<1100$) distortions in the framework of the standard LCDM model of cosmology arises via   Compton scattering of the CMB photons
 \cite{1975SvA....18..413I} with the hot gas in the intracluster medium (ICM) and intergalactic medium (IGM) \cite{PhysRevD.49.648, PhysRevD.61.123001,Zhang:2004fh,Hill:2015tqa}.  At  early times ($z>1100$), dissipation of primordial acoustic waves \cite{1970Ap&SS...7....3S,1991ApJ...371...14D, 2012ApJ...758...76C} and the energy injection due to recombinations of hydrogen and helium \cite{1968ZhETF..55..278Z, 2009AN....330..657S} are the primary sources of spectral distortions in the CMB according to the standard model of cosmology. Along with these effects, spectral distortions in the microwave sky can also originate from axions \cite{Mukherjee:2018oeb, Mukherjee:2018zzg, Mukherjee:2019dsu}, decay of particles \cite{PhysRevLett.70.2661, Chluba:2013pya}, primordial black holes \cite{Carr:2009jm},  and small-scale magnetic fields \cite{PhysRevLett.85.700}. These distortions can be variously classified on the basis of their spectral shapes as $y$-type, $\mu$-type, $i$-type (or residual $r$-type) \cite{Khatri:2012tw}, relativistic SZ \cite{Hill:2015tqa}, recombination lines \cite{1968ZhETF..55..278Z, 2009AN....330..657S} and  $\alpha$-type (axions) \cite{Mukherjee:2018oeb, Mukherjee:2018zzg, Mukherjee:2019dsu}.

 {Injection of energy at different epochs of the  early Universe leads to several of these spectral distortions. Injection of energy at $z \geq 2 \times 10^6$ does not lead to any spectral distortion in the CMB blackbody spectrum  as the interactions between the charged particles and photons due to bremsstrahlung and double Compton emission are efficient. Energy injection at a later epoch  $5\times 10^4\leq z\leq 2 \times 10^{6}$, produces $\mu$-type spectral distortion when Compton scattering distributes the injected energy, creating a Bose-Einstein distribution with a chemical potential $\mu$. For $z\leq 5\times 10^4$, the energy injection leads to $y$-distortions in the CMB blackbody. As a result, measurement of the  spectral distortion signal is crucial for understanding the thermal history of the Universe.}
From the various sources of spectral distortions, we expect the strength of the $y$-type and $\mu$-type distortions to be $2\times \, 10^{-6}$ \cite{Hill:2015tqa} and $2\times \, 10^{-8}$ \cite{Chluba:2012gq} respectively. FIRAS 
 placed  observational upper bounds on the absolute value of $y$-type and $\mu$-type distortions at $2.5 \times 10^{-5}$ and $3.3 \times 10^{-4}$ at $95\%$ confidence level \cite{1994ApJ...420..439M}.  
 A CMB mission Primordial Inflation Explorer (PIXIE)  \cite{2011JCAP...07..025K, 2014AAS...22343901K} was recently reproposed to measure the spectrum of the microwave sky using a Fourier-transform spectrometer (FTS) with $400$ frequency channels between $30$- $6000$ GHz, able to improve on FIRAS by some four orders of magnitude but was not approved.  { A recent analysis  \cite{Abitbol:2017vwa} has explored the requirements of an FTS to detect different spectral distortion signals in the presence of foreground contamination.}
 
\begin{figure}[h]
\centering
\begin{subfigure}{0.5\textwidth}
\centering
\includegraphics[trim={0 0 0 0cm}, clip, width=1\textwidth]{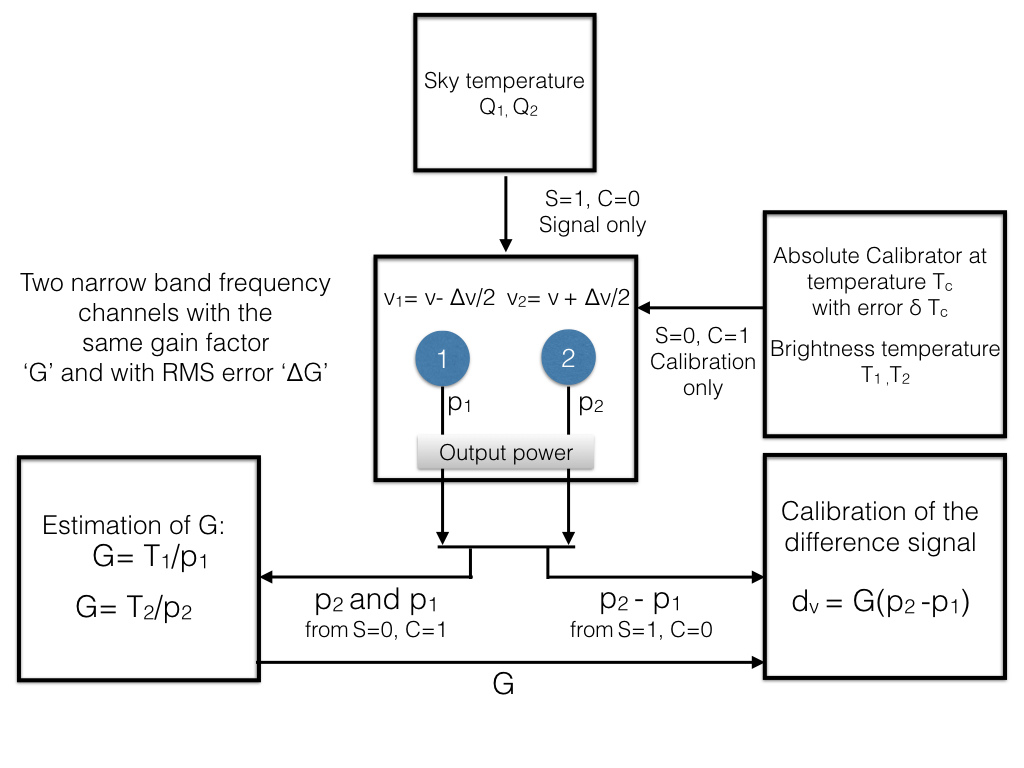}
 \caption{}\label{Fig:sc2}
     \end{subfigure}
     \captionsetup{singlelinecheck=on,justification=raggedright}
 \caption{A schemstic diagram showing the working principle of the inter-frequency differential  (IFD) technique for a single pair of frequency channel working at a central frequency $\nu$ with two frequency channels $\nu_1$ and $\nu_2$. The absolute calibrator known at a temperature $T_c$ with a calibration error $\delta T_c$. $S=1, \,C=0$ and $S=0,\, C=1$ indicates two modes of operation when the sky signal and  absolute calibrator are shown to the detectors respectively. }\label{Fig:sc-2}
\end{figure}
We propose here a new method using an Inter-Frequency Differential (IFD) technique to measure the spectrum of the microwave sky \textit{without using a spectrometer}.  
CMB spectral distortions of known spectral shapes can be measured from the differential measurement of the sky intensity between two frequency channels at unprecedented sensitivity by use of an imaging telescope along with an onboard inter-frequency calibrator \footnote{ {Calibration of the difference in  intensity between a pair of frequency channels with a calibrator of known spectrum.}}. 
{A schematic diagram of this concept is shown in Fig. \ref{Fig:sc-2}. We propose that in a particular sky direction, we can use the differences in the sky intensity in  multiple pairs of frequency channels as an observable to characterize different types of spectral distortions and foreground contamination.}

To study the feasibility of this concept in the presence of astrophysical foregrounds, we apply a Bayesian MCMC  {setup} \textbf{InCAS-MC} (Instrument optimization for Cosmological and Astrophysical Signal- MCMC) to explore different combinations of instrument parameters such as the number of frequency channels $N_{c}$, the width of each channel $\Delta \nu$,and  the relative accuracy of the   calibrator $\Delta_g$ and instrument noise $\Delta_N$. 
For the joint estimation of six parameters (three cosmological and three foreground), we find that the  inter-frequency differential (IFD) technique is capable of measuring $y$ and $\mu$ distortions at SNR $>3$ with a number of frequency channels $N_{c}$ between $30$ to $100$ with an inter-frequency   calibrator error $\Delta_g \leq 10^{-9}$ and instrument noise $\Delta_N \leq 0.1$ Jy/sr. For a higher value of the instrument noise (or calibration error), the measurement of $\mu= 2\times 10^{-8}$ becomes impossible, but a high SNR measurement of $y= 2\times 10^{-6}$ is feasible.  \text{InCAS-MC} is applicable to any other foreground models to find the best instrument configuration of an imager required to measure spectral distortion signals. 

\section{\textbf{Formalism and set-up of the IFD technique}}
The central idea of the IFD technique is to use the difference in the sky intensity between a pair of frequency channels \cite{Mukherjee:2018fxd}. 
\subsection{\textbf{Measurement and calibration}} The total power measured by a bolometric detector in  sky direction $\hat x$ over a given frequency band can be written as \cite{1994JAP....76....1R}
\begin{equation}\label{bolo-1}
\begin{split}
p_i(\hat x)= \int_{0}^{\infty} d\nu\, w(\nu_i, \Delta \nu) I_\nu A\Omega\,,
\end{split}
\end{equation}
 where $I_\nu$ is the intensity at frequency $\nu$, $A\Omega$ is the etendue which is $(c/\nu)^2$ for a diffraction-limited CMB experiment, and $w(\nu_i, \nu)$ is the transmission function, which we have chosen as
\begin{align}\label{weight-1}
\begin{split}
w(\nu_i, \Delta \nu) &= 1\, \text{for}\, \nu_i -\Delta \nu/2 <\nu<\nu_i +\Delta \nu/2,\\
&= 0,\, \text{otherwise}.
\end{split}
\end{align}
The measured power in each frequency channel is then subtracted from the adjacent frequency channel to obtain the  derivative signal (inter-frequency differential signal) which we define as
$\Delta p_{i}(\hat x)= p_{i+1}(\hat x) - p_i(\hat x).$
The derivative signal $\Delta p_{i}$ carries the information of the difference in the power between two frequency channels with central frequency $\nu_{i+1}$ and $\nu_i$. So for a  total number of frequency channels $N_{c}$, we have $N_{c}/2$ independent differential signals.

In order to convert this observed power into a meaningful sky signal, we need to calibrate the observed derivative signal with a known   inter-frequency calibrator. As shown in Fig. \ref{Fig:sc-2}, during the calibration mode, the gain factor $G_i$ needs to be calibrated for each frequency channels with an absolute calibrator of known temperature $T_c$. The corresponding RMS error in the gain factor $\Delta G$ is going to be related to the error in the calibrator $\delta T_c$ and the error in the output power $\delta p$ by the relation
\begin{align}\label{calib-1b}
\begin{split}
\Delta_g \equiv \frac{\Delta G_i}{G} = \sqrt{\bigg(\frac{\delta T_c}{T_c}\bigg)^2 + \bigg(\frac{\delta p_i}{p_i}\bigg)^2}.
\end{split}
\end{align}
One of the requirements of the IFD technique is to have the same gain factor $G_i= G_{i+1}$ for the pair of frequency channels which are used to calculate the derivative signal, in order to reduce the contamination from the calibration error. The corresponding calibrated derivative signal can be written in terms of $\Delta p_i$ as
\begin{align}\label{calib-1}
\begin{split}
d_{i}(\hat x)= G\Delta p_{i}(\hat x) + \Delta_g G\Delta p_{i}(\hat x). 
\end{split}
\end{align} 
If the condition $G_i=G_{i+1}$ cannot be satisfied from an instrument design, then the difference of the calibrated signal needs to be estimated to apply the IFD technique. In this case, the calibrated derivative signal becomes 
\begin{align}\label{calib-1a}
\begin{split}
d_{i}(\hat x)=& G_{i+1} p_{i+1}(\hat x) - G_{i}p_{i}(\hat x) \\ &+ \Delta_{g_{i+1}} G_{i+1} p_{i+1}(\hat x) - \Delta_{g_{i}} G_{i} p_{i}(\hat x). 
\end{split}
\end{align}

In this analysis, we consider the case with the same gain factor $G_{i+1}= G_i$ for the pair of frequency channels which are used to obtain the derivative signal and use Eq.\eqref{calib-1}. If a future mission is unable to satisfy this criterion, then the contribution to the covariance matrix from the systematic uncertainty is going to increase.   By using Eq. \eqref{calib-1a} instead of Eq. \eqref{calib-1} in the IFD setup, we estimate the change in the variance for a fixed instrument noise $\Delta_N$ and calibration error $\Delta_g$. The change in the variance depends on the central value of the frequency channel $\nu_c$ and also on the number of  frequency channels $N_c$. For the choices of instrument design suitable for making an unbiased estimation of the spectral distortion signal (see Sec. \ref{resultsin} and Sec. \ref{foregrounds}), we find that the maximum increase in the variance is about $23\%$ at the frequency channel $\nu_c=1000$ GHz for the case with $N_c=100$. The effect on the variance reduces with the reduction in the value of both $\nu_c$ and $N_c$.

The calibrated derivative signal $d_{i}$ can be expressed as a combination of the sky signal $\Delta s_{i}$ and instrument noise $n_{i}$ as
\begin{align}\label{skymodel-1}
\begin{split}
d_{i}(\hat x)= \Delta s_{i}(\hat x) + \Delta_g \Delta s_{i}(\hat x)+ n_i(\hat x) + n_{i+1}(\hat x).
\end{split}
\end{align}
Here, the contribution from the instrument noise \cite{Mather:82} arises from both the channels which are used to obtain the differential signals. The sky-averaged derivative signal $d_{i}= \int d^2{\hat x} d_{i}(\hat x)/ 4\pi$ can be written as
\begin{align}\label{skymodel-2}
\begin{split}
d_{i}= \Delta s_{i} + \Delta_g\Delta s_{i},
\end{split}
\end{align}
where  $\Delta s_{i} $ is the all-sky average differential signal and the noise terms are expected to have zero mean $\int d^2{\hat x}\,\, n_{i+1}(\hat x)=0$. The calibration error acts as a source of contamination to the actual sky signal and hence can bias the inferred value of the signal. The corresponding noise covariance matrix $\Sigma_{ij}$ for the derivative sky signal can be written as 
 \begin{align}\label{covmat-1}
\begin{split}
\Sigma_{ij}= \bigg((\Delta_g)^2 \Delta s_i^2 +\mathcal{N}_i + \mathcal{N}_{i+1}\bigg) \delta_{ij},
\end{split}
\end{align}
where $\left<n_{i}(\hat x) n_{j}(\hat x) \right>= \mathcal{N}_i \delta_{ij}$ is the instrument noise power spectrum. $\mathcal{N}_i$ can be related to the corresponding intensity noise by the relation $(\Delta_N)_i= \sqrt{\mathcal{N}_i}/(A\Omega \Delta \nu (\alpha\epsilon f))$, where $\alpha$ is the detector absorptivity, $\epsilon$ is the source emissivity and $f$ is the transmissivity of the optical system. 
Each frequency channel is only used once to construct the derivative signal. As a result, we do not introduce any correlation between any pair of the derivative signals and we can expect the covariance matrix to be diagonal.  {Multiple use of the same frequency channels to construct all possible difference signal can lead up to  $N_c(N_c-1)/2$ combinations with non-zero  off-diagonal terms in the covariance matrix. The possibility of accessing all the combinations of the differential signal requires a stable gain calibration for all the pairs, which may become difficult to achieve for a large number of frequency channels over a wide frequency range. In principle, one can implement the IFD method for all frequency channels and use Eq. \eqref{calib-1a} for the calibration. We will examine this aspect in a future analysis.} 

\subsection{\textbf{Sky Signal: spectral distortion signals and foreground models}} The theoretical model of the {IFD} sky signal $\Delta s_i$  between a pair of frequency channels can be written as 
\begin{align}\label{skymodel-3}
\begin{split}
\Delta s_i (\hat x) = \mathcal{I}_{i+1} (\hat x) -\mathcal{I}_{i} (\hat x),
\end{split}
\end{align}
where $\mathcal{I}_{i}$ can be written in terms of the intensity of different components as \cite{Adam:2015wua}
\begin{align}\label{skymodel-4}
\begin{split}
\mathcal{I}_{i}=& \int d\nu\, w(\nu_i, \Delta \nu)\, A \Omega  \bigg(I_{BB|T_{CMB}}(\nu)  + \frac{\Delta T}{T_{CMB}} I_{\Delta T} + \mu I_{\mu}(\nu) \\& + y I_{y}(\nu)   + I_{other}(\nu) + A_{sync}  I_{sync}(\nu)+ A_{Dust} I_{Dust}(\nu) \\& + A_{CIB} I_{CIB}(\nu) + A_{sd} I_{sd}(\nu)+ A_{ff} I_{ff}(\nu)\bigg),
\end{split}
\end{align}
where $I_{BB|T
_{CMB}}$ denotes the intensity of the CMB blackbody for $T_{CMB}= 2.725$ K and $ I_{\Delta T}$ is the intensity due to the unknown part of the CMB sky temperature $ \Delta T=  T^{true}_{CMB} - T_{CMB}$.   $I_{\mu}$ and  $I_{y}$ denoted the distortion due to $\mu$ and $y$-distortion.  $I_{other}$ denotes the distortion due to several other distortions due to relativistic SZ, axion and $i$-type distortion.  The astrophysical signals such as synchrotron ($I_{sync}$) radiation, warm dust emission ($I_{Dust}$), the cosmic infrared background  ($I_{CIB}$), spinning-dust ($I_{sd}$), and  free-free ($I_{ff}$) also emit brightly in these frequency bands and hence act as a potential sources of  contamination for the cosmological signals. The frequency dependence of the sky intensity for different cosmological components is well known from the underlying theory. This can be written as
\begin{align}\label{source-1a-int}
\begin{split}
I_{\text{BB}}({\nu}) &= \frac{2(kT_{CMB})^3}{c^2h^2}\frac{{x^3}}{(e^x-1)},\\
I_{\Delta T}(\nu)&= \frac{2(kT_{CMB})^3}{c^2h^2}\frac{{x^4}e^x}{(e^x-1)^2},\\
I_\mu(\nu)&= \frac{2(kT_{CMB})^3}{c^2h^2}\frac{{x^3}e^x}{(e^x-1)^2}\bigg(\frac{x}{2.1923}-1\bigg),\\
I_y(\nu)&= \frac{2(kT_{CMB})^3}{c^2h^2}\frac{{x^4}e^x}{(e^x-1)^2}\bigg(x\coth{\big(x/2\big)}-4\bigg),
\end{split}
\end{align}
where $x= h\nu/kT_{CMB}$ and $k$ is the Boltzmann constant. The intensity of the foregrounds is not well known and depends on several model parameters related to the physics of the interstellar medium. The intensity of the foregrounds can be written in terms of the currently best known value from the CMB experiments such as WMAP \cite{2013ApJS..208...19H} and Planck \cite{2016A&A...594A...1P} as
\begin{figure*}[t]
\centering
     \includegraphics[trim={0.5cm 0.2cm 0.5cm 0.8cm}, clip, width=0.9\textwidth]{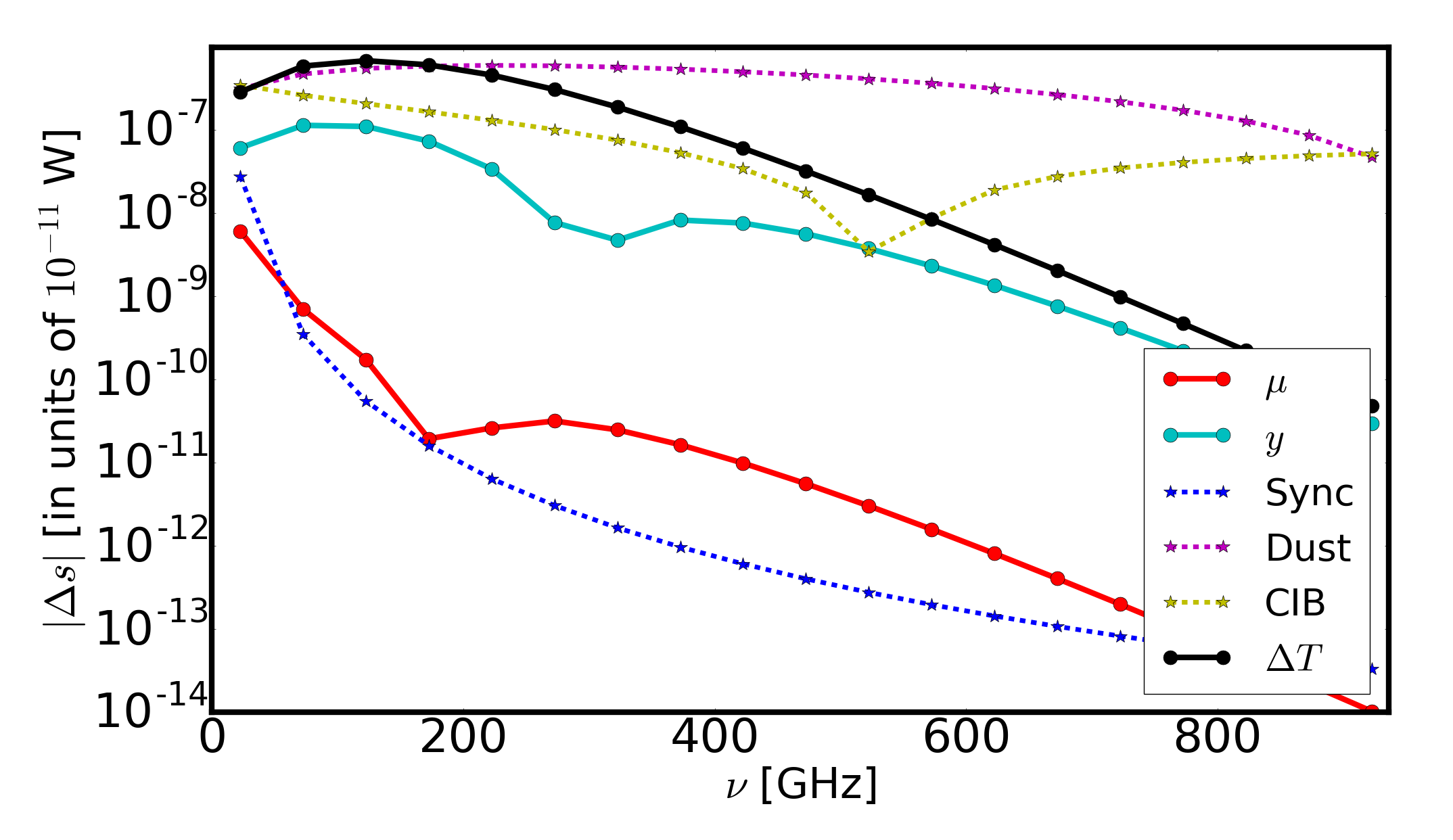}
     \captionsetup{singlelinecheck=on,justification=raggedright}
 \caption{ {IFD spectrum of  cosmological signals and astrophysical foregrounds for $N_{c}=40$ and  $\Delta \nu= 25$ GHz for fiducial values $\mu= 2\times 10^{-8}$, $y= 2\times 10^{-6}$, $\Delta T= 10^{-4}$ K, $A_{sync}= 288$ Jy/sr, $A_{Dust}= 1.3$ MJy/sr, $A_{CIB}= 0.35$ MJy/sr.}}\label{Fig:sc-1}
\end{figure*}
\begin{align}\label{source-2a-int}
\begin{split}
I_{\text{Dust}}({\nu})&= \bigg(\frac{x_D^{\beta_D+{3}}}{e^{x_D}-1}\bigg), x_D= \frac{h\nu}{kT_D}, T_D=21 \text{K}, \beta_D=1.55,\\
I_{\text{CIB}}({\nu})&= \bigg(\frac{x_{C}^{\beta_C+{3}}}{e^{x_{C}}-1}\bigg), x_{C}= \frac{h\nu}{kT_{C}}, T_{C}=18.8 \text{K}, \beta_{C}=0.86,\\
I_{\text{sync}}({\nu})&=  \bigg(\frac{\nu_0}{\nu}\bigg)^{\alpha_{sync}} \hspace{0.5cm} \nu_0= 100 \text{GHz},  \alpha_{sync}=0.82,\\
I_{\text{ff}}(\nu)&=  \nu^2T_e(1-e^{-\tau}), \hspace{0.5cm} \\&\tau= 0.05468 T_e^{-3/2}\nu_9^{-2} \log(e^{[5.96- \sqrt{3}{\pi}log(\nu_9T_4^{-3/2})]}+e), \\& 
\nu_9= \frac{\nu}{GHz}, T_4= T_e/10^4,\\
I_{\text{sd}}({\nu})&= \frac{f_{sd}(\nu.\nu_{p0}/\nu_p)}{f_{sd}(\nu_0.\nu_{p0}/\nu_p)}, \nu_{p0}=30 \text{GHz}, f_{sd}= \text{templates}.
\end{split}
\end{align}

{The IFD spectra for different cosmological signals and  foreground components are shown in Fig. \ref{Fig:sc-1} for $N_c=40$ and $\Delta \nu= 25$ GHz using Eq. \eqref{skymodel-1}. {We have taken the fiducial values of the foreground parameters as  $A_{sync}= 288$ Jy/sr, $A_{Dust}= 1.3$ MJy/sr and $A_{CIB}= 0.35$ MJy/sr \cite{Abitbol:2017vwa}.}

\section{\textbf{Likelihood Analysis}}
\begin{figure*}
\centering
     \includegraphics[trim={0 0 0 0cm}, clip, width=1.\textwidth]{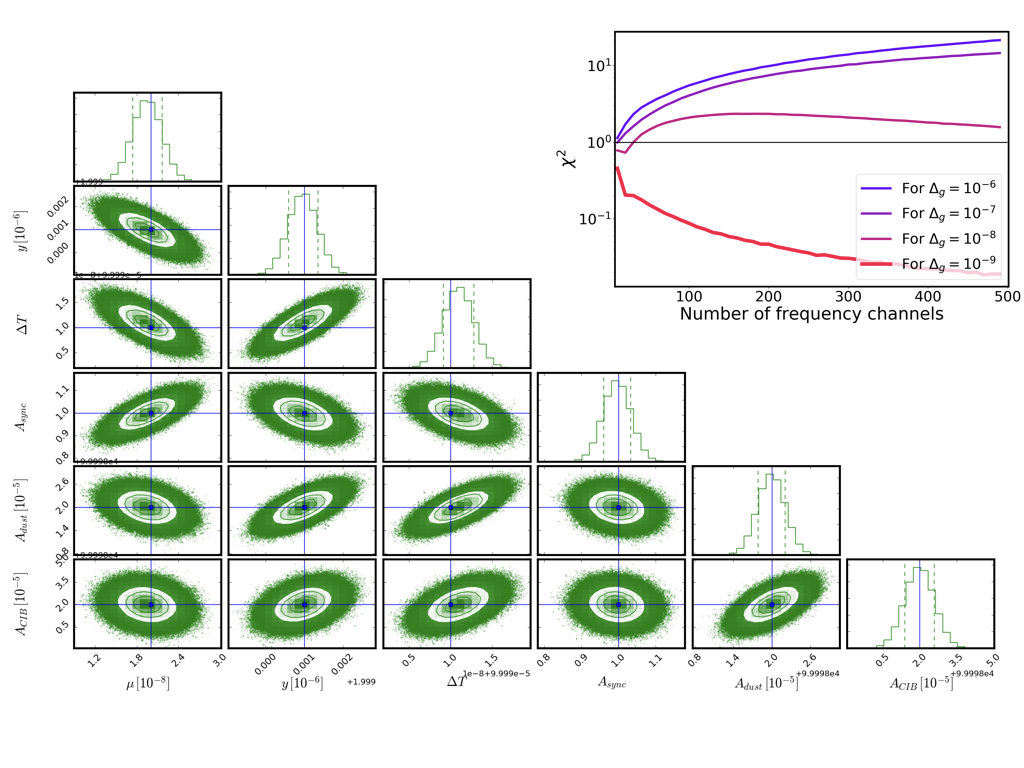}
     \captionsetup{justification=raggedright}
 \caption{Triangle plot: The posterior distribution of the cosmological and astrophysical foregrounds obtained using the framework of InCAS-MC for $\nu_{min}=10$ GHz, $\nu_{max}= 1010$ GHz, $N_c=40$, $\Delta_g= 10^{-9}$ and $\Delta_N= 0.1$ Jy/sr.  {The blue line denotes the injected value of the signals $\mu= 2\times 10^{-8}$, $y= 2\times 10^{-6}$, $\Delta T= 10^{-4}$ K, $A_{sync}= 1$ (in units of $288$ Jy/sr), $A_{Dust}= 1$ (in units of $1.3$ MJy/sr), $A_{CIB}= 1$ (in units of $0.35$ MJy/sr) used in the analysis and the contours are drawn for $1\sigma$, $1.5\sigma$ and $2\sigma$.} In the upper right corner: we plot the value of $\chi^2$ defined in Eq. \eqref{gamma} for $\Delta_N=0.1$ Jy/sr (in thick red line). We have also plotted the value of the $\chi^2$ for other values of $\Delta_N$.}\label{Fig:T-1}
\end{figure*}
The feasibility of the IFD method to measure the spectral distortion depends upon how robustly this  technique can estimate spectral distortion signals in the presence of other foreground contaminations. The efficiency of separating the spectral distortion signal from other foregrounds  depends upon the choice of instrument parameters such as the smallest frequency channel $\nu_{min}$, the largest frequency channel $\nu_{max}$, the bandwidth of each channel $\Delta \nu$, the number of frequency channels $N_{c}$,   calibrator noise $\Delta_g$ and instrumental noise $\Delta_N$. Hence we need  to optimize the instrumental configuration.

We develop a Markov Chain Monte Carlo (MCMC)  {setup} InCAS-MC (Instrument optimization for Cosmological and Astrophysical Signal-MCMC) under a Bayesian framework to find the feasible instrument configurations which can make  unbiased and high SNR measurements of the spectral distortion signals in the presence of astrophysical foregrounds. InCAS-MC can be applied to more general foreground models to optimize instrumental configuration for a desired cosmological signal.  For a particular choice of instrument configuration, we obtain the joint posterior distribution $\mathcal{P} (\mathbf{ \Theta}| \mathbf{d}, \mathbf{\Phi})$ of the spectral distortion and foreground parameters in terms of the likelihood  $L(\mathbf{d}|\mathbf{ \Theta}, \mathbf{\Phi})$ and prior $\Pi(\mathbf{ \Theta})$ by using  Bayes' theorem \cite{bayes}  
$\mathcal{P} (\mathbf{ \Theta}| \mathbf{d}, \mathbf{\Phi})= L(\mathbf{d}|\mathbf{ \Theta}, \mathbf{\Phi})\Pi(\mathbf{ \Theta}),
$
where $\mathbf{ \Theta}$ is the set of parameters related to spectral distortion and astrophysical foreground, $\mathbf{\Phi}$ denotes the set of instrument parameters such as $\nu_{min}, \nu_{max}, \Delta \nu, N_{c}, \Delta_N, \Delta_g$.

We can express the  likelihood, assumed to be  Gaussian, in terms of the data $\mathbf{d}$ and differential sky model $\mathbf{\Delta s}$ as 
\begin{equation}\label{likelihood}
L(\mathbf{d}|\mathbf{\Theta}, \mathbf{\Phi})= \frac{\exp\bigg({-\frac12 (\mathbf{d}(\mathbf{\Phi}) -\mathbf{\Delta s})\mathbf{\Sigma}^{-1}(\mathbf{d} (\mathbf{\Phi})-\mathbf{\Delta s})^\dagger\bigg)}}{\sqrt{(2\pi)^{N_{c}/2}|\Sigma|}}, 
\end{equation}
where $\mathbf{\Sigma}$ is the  covariance matrix mentioned in Eq. \eqref{covmat-1} and depends on $\mathbf{\Phi}=\{N_{c}, \Delta_g, \Delta_N. \Delta \nu, \nu_{min}, \nu_{max}\}$.

 \section{\textbf{Results using InCAS-MC for six parameters}}\label{resultsin}
We have taken six cosmological and astrophysical foreground parameters  $\mathbf{\Theta}= \{\mu, y, \Delta T, A_{sync}, A_{Dust}, A_{CIB}\}$ with a flat prior.  {Instrumental parameters such as number of frequency channels $N_{c}$ are varied from $10$--$500$ in steps of ten, and $\Delta_g$ and $\Delta_N$ are varied by a factor of ten from  $10^{-6}$--$10^{-9}$ and $10$--$0.1$ Jy/sr, respectively.}  
There are in all  {$600\,=(50\times4\times3)$} instrument configurations for which we have performed the MCMC analysis.  {Other instrument parameters such as $\nu_{min}$ and $\nu_{max}$ are kept fixed at $10$ GHz and  $\nu_{max}= 1010$ GHz respectively}. $\Delta \nu$ is related to these  parameters by  {the relation} $\Delta \nu= (\nu_{max}- \nu_{min})/N_{c}$.

\begin{table*}[t]
\centering
\begin{tabular}{|c |c   c  c  c  c| c   c  c  c c| c   c  c  c c|}
\hline 
$\Delta_N$ (Jy/sr) &\multicolumn{5}{|c|} {$\Delta_g= 10^{-7}$} & \multicolumn{5}{|c|} {$\Delta_g= 10^{-8}$} & \multicolumn{5}{|c|} {$\Delta_g= 10^{-9}$} \\
\hline \hline

 -& $N_{c}$& $b_\mu$&$\sigma_\mu$&$b_y$& $\sigma_y$& $N_{c}$& $b_\mu$&$\sigma_\mu$&$b_y$& $\sigma_y$& $N_{c}$& $b_\mu$&$\sigma_\mu$&$b_y$& $\sigma_y$\\   \hline 
 10 & 20&125&102&13.2&34&20&131&102&9&33&30&127&102&16.3&26.3     \\ \hline
 1&110& 3.7& 18.9&8.9&6.7 &40& 3.6 &15.6&1.44&3.4 &40& 5.3& 15.6&0.79&3.3 \\ \hline
0.1& 480 &13.9& 4.8&5.7&2.1& 120 &3.4& 3.5&0.46&0.76 &40& 0.5& 2.1&0.03&0.38  \\ \hline
\end{tabular}
\captionof{table}{Minimum error-bar $\sigma$ and the  corresponding bias $b$  {(both in the units of $10^{-9}$)} in the inferred value of $\mu$ and $y$ distortions and number of frequency channels $N_c$ for different values of instrument noise $\Delta_N$ and calibration error $\Delta_g$ obtained using InCAS-MC.} \label{tab:1}
\end{table*}
We use the publicly available MCMC sampler emcee: MCMC Hammer \footnote{\url{http://dfm.io/emcee/current/}} \cite{2013PASP..125..306F} to sample the distribution of $\mathbf{\Theta}$. We have chosen number of walkers ($N_{walk}$=1000) \footnote{{$N_{walk}$ is a parameter of the emcee code related to the chains of the Metropolis-Hastings algorithm.}} 
and number of steps $n_{step}= 10^4$,  discarding the first $2\times 10^3$ steps as the \lq burn-in\rq. The joint posterior distribution of the $\Theta_i$ for one of the best instrument configurations $N_{c}=  40$, $\Delta_g= 10^{-9}$ and $\Delta_N= 0.1$ Jy/sr for only six parameters  is shown in Fig. \ref{Fig:T-1}. 
 
\begin{figure*}
\centering
     \includegraphics[trim={0 0 0 0cm}, clip, width=1.\textwidth]{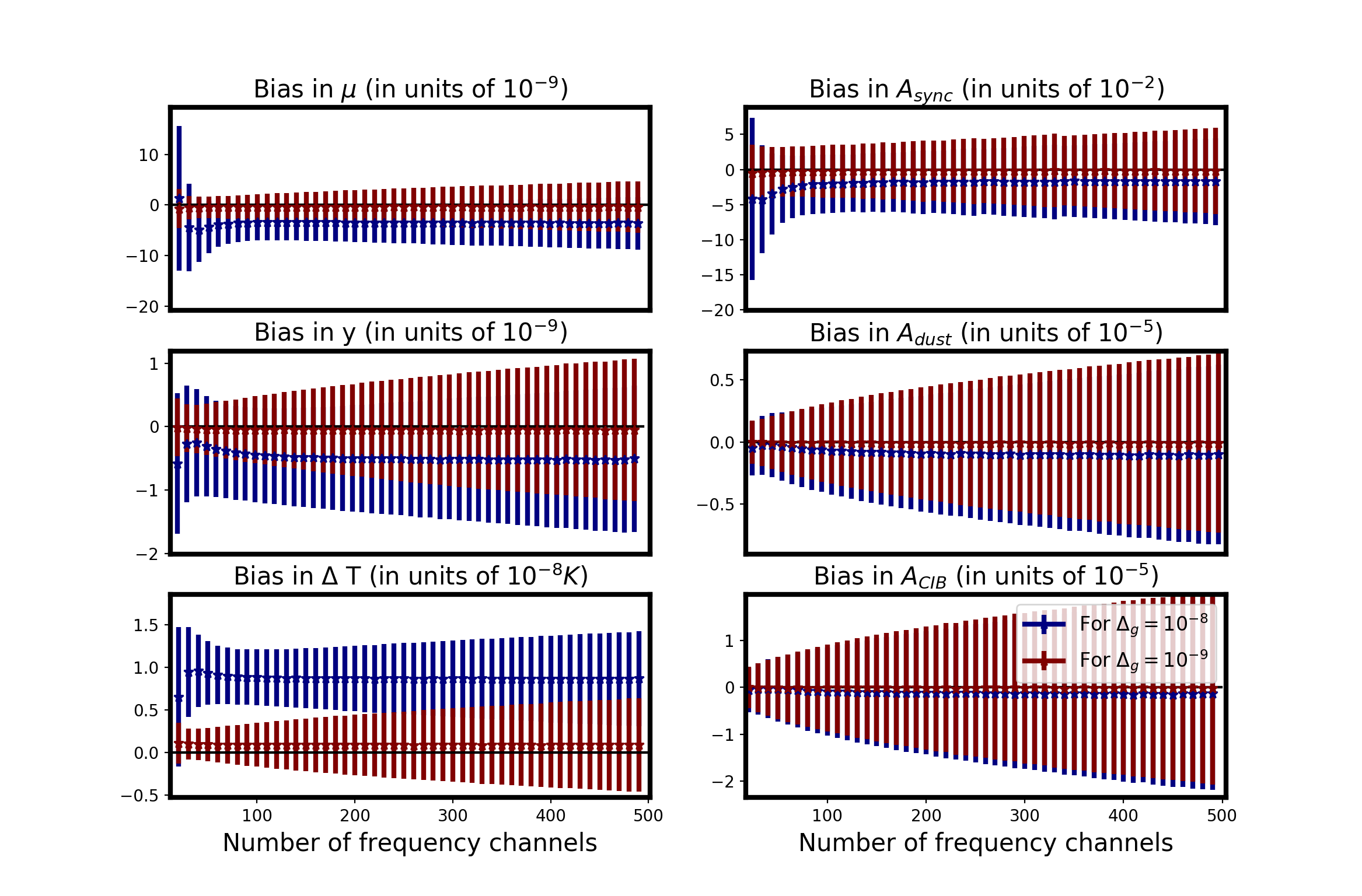}
     \captionsetup{singlelinecheck=on,justification=raggedright}
 \caption{Bias and error bars in  cosmological and astrophysical foreground signals for different numbers of frequency channels $N_c\geq 20$ and  calibration errors $\Delta_g= 10^{-8}$, $\Delta_g= 10^{-9}$, for $\Delta_N=0.1$$\rm {Jy/sr},$ $\nu_{min}=10$ GHz, $\nu_{max}=1010$ GHz obtained using InCAS-MC. }\label{Fig:8_9_c}
\end{figure*}

The posterior distributions in Fig. \ref{Fig:T-1} indicate that we can make reliable detections of the $\mu=2\times 10^{-8}$, $y= 2 \times 10^{-6}$ distortion with an SNR of $10$ and $5 \times 10^3$  {respectively}. The mean of the posterior distribution recovers the injected value successfully and the bias in the mean value is within $1$--$\sigma$. {We define a criterion to evaluate the choice of the instrument parameters }
\begin{equation}\label{gamma}
\centering
\chi^2 = \frac{1}{N_{p}}\sum^{N_p}_{i,j=1}(\hat \theta_i - \bar\theta_i)C^{-1}_{\theta_{i}\theta_{j}}(\hat \theta_j - \bar\theta_j),
\end{equation}
where the sum runs over all the parameters ($N_p=6$   in} this analysis), $ \hat \theta_i$ is the mean value obtained from the posterior distribution, $ \bar \theta_i$ defines the true value of the parameter and $C^{-1}_{\theta_{i}\theta_{j}}$ is the inverse of the covariance matrix obtained from the MCMC samples.  {This quantity can be chosen  between $[\chi^2_{min},1]$ for an optimized choice of instrument parameters.} The variation of $\chi^2$ with $N_c$ and $\Delta_g$ are also shown in Fig. \ref{Fig:T-1}.  {This clearly indicates that $\Delta_g \leq 10^{-9}$ is the best scenario to measure $\mu=2 \times 10^{-8}$ using the IFD method}. 

The variation of the bias (mean of the posterior- injected signal) and error-bar on the cosmological and astrophysical parameters for different choices of instrument configurations are shown in Fig. \ref{Fig:8_9_c}. 
We find from this analysis that a large number of frequency channels (or narrow frequency bandwidth $\Delta \nu$) can lead to large error bars. This is mainly due to two reasons. Firstly, the inter-frequency differential signal starts appearing to be smoother and less informative as one goes to very narrow frequency bands. Secondly, with the decrease in the value of  $\Delta \nu$, the total amount of power in each frequency band reduces and makes the differential signal very small.  

\begin{figure*}[t!]
    \centering
    \includegraphics[trim={0 0 0 0cm}, clip, width=0.8\textwidth]{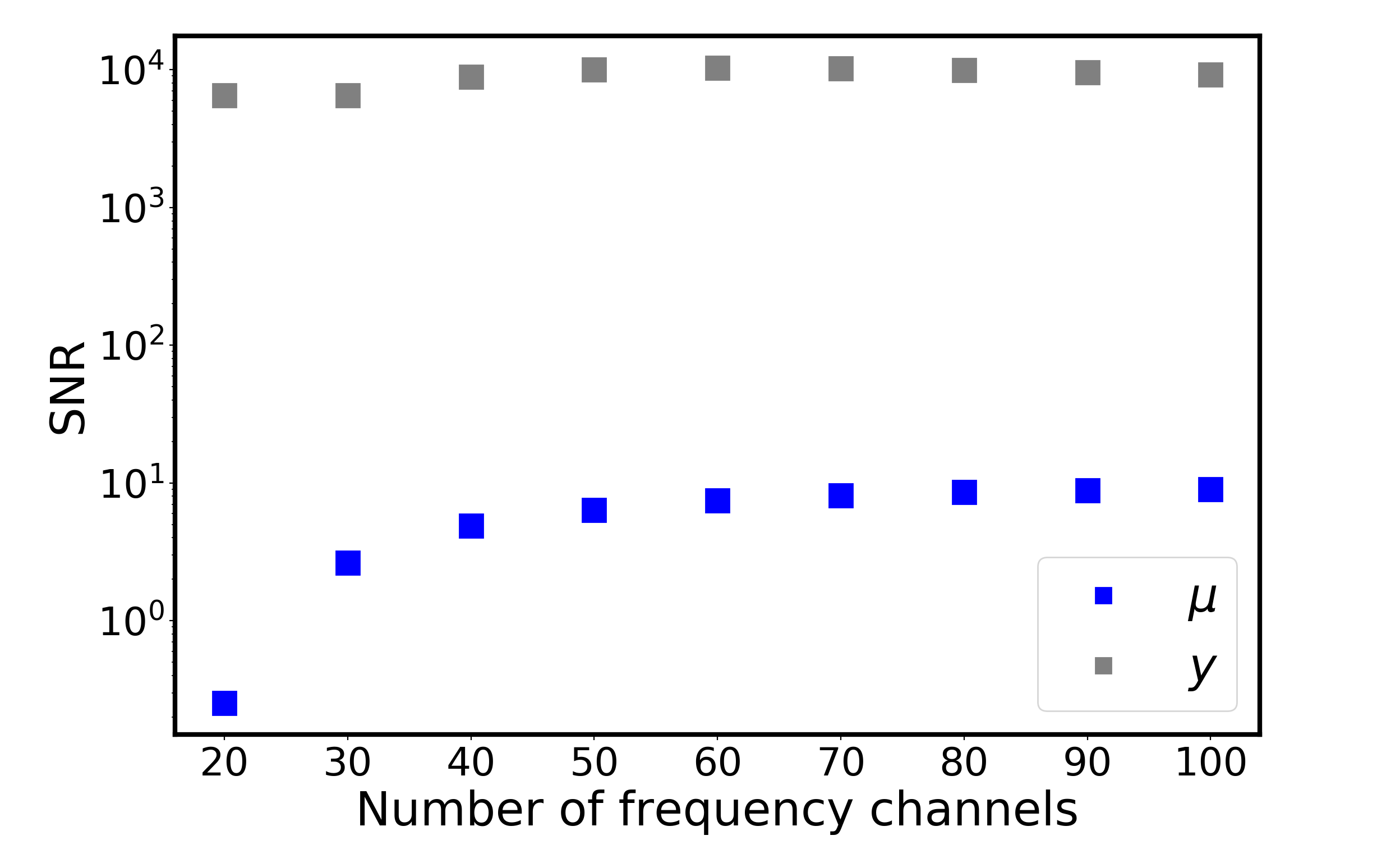}
    \caption{The signal-to-noise ratio (SNR)($\equiv  (\mu=2\times 10^{-8})/\sigma_{\mu}$ and $(y=2\times 10^{-6})/\sigma_y$) for $\mu$ and $y$ distortions are shown as a function of the number of frequency channels $N_c$ for the instrument noise  $\Delta_N= 0.01$ Jy/sr and calibration noise $\Delta_g= 10^{-9}$ after marginalizing over all the parameters with a $10\%$ prior on the amplitude of the synchrotron emission $A_{sync}$.}
    \label{fig:SNR_fisher}
\end{figure*}

 For the foreground parameters, we find $\Delta_g= 10^{-9}$, $\Delta_N=0.1$ Jy/sr and $20\leq N_{c} \leq 100$ provides the best scenarios to detect the spectral distortion signals, as can be seen from red lines in Fig. \ref{Fig:8_9_c}.  We summarize the performance of different instrument configurations to detect $\mu$ and $y$ distortion signal in Table \ref{tab:1} by quantifying the bias $b$ in the mean value of the posterior with respect to the true value and  {the corresponding error-bar $\sigma$}. The instrument configurations with minimum bias are considered as the best choices to measure spectral distortion of CMB using the IFD technique.

\section{Fisher analysis with eight foreground parameters and three spectral distortion parameters}\label{foregrounds}
Our InCAS-MC analysis suggests that the posterior shapes are very close to Gaussian. We can exploit this fact to  explore forecasts that include richer foreground models with more parameters using a Fisher information matrix approach. In order to show the ability of the IFD technique to distinguish between the spectral distortion signal and the astrophysical foregrounds, we have considered eight foreground parameters such as $A_{sync}$, $A_{Dust}$, $A_{CIB}$, $\alpha_{sync}$, $\beta_{D}$, $\beta_{C}$, $T_{D}$, $T_{C}$ along with three cosmological parameters such as $\mu$, $y$ and $\Delta T$ to make a Fisher estimate of the error-bar on the cosmological parameters using the relation
\begin{equation}
    F_{ij}= -\bigg\langle \frac{\partial^2\mathcal{L}}{\partial p_i \partial p_j}\bigg\rangle,
\end{equation}
where, $p_i$ denotes eight foreground parameters and three cosmological parameters and $\mathcal{L}$ denotes the log likelihood $\mathcal{L}= \ln{L}$, where the likelihood is given in Eq. \eqref{likelihood}. The Fisher estimation is applied for the case with large number of parameters, in order to make  efficient estimates of the error bars for different instrument configurations. 

\begin{figure*}
\centering
     \includegraphics[trim={1.5cm 1.cm 3.cm 2.cm}, clip, width=1.\textwidth]{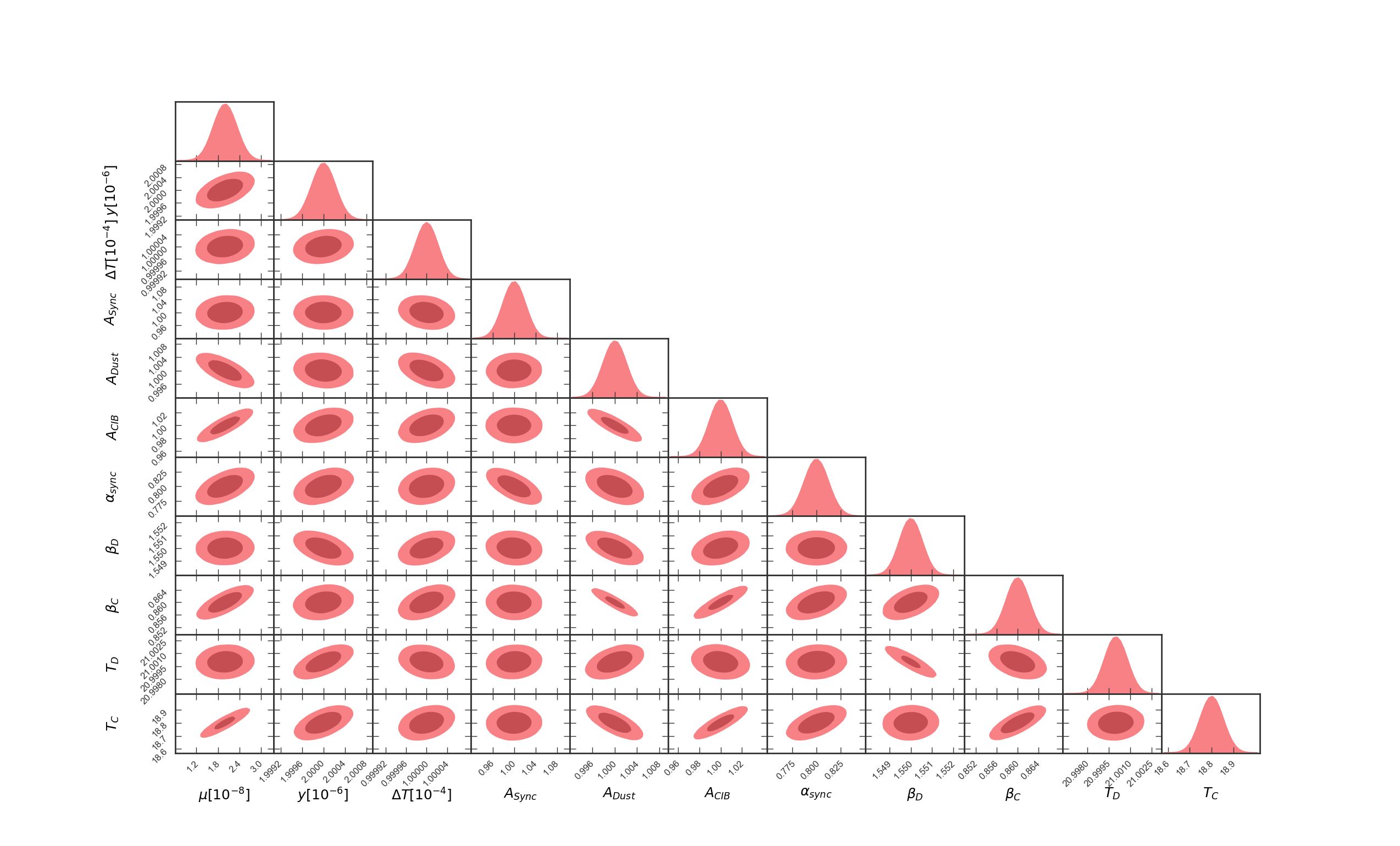}
     \captionsetup{justification=raggedright}
 \caption{Triangle plot: The $1$-$\sigma$ and $2$-$\sigma$ contours  using the Fisher analysis with a $10\%$ prior on the amplitude of the synchrotron parameter $A_{sync}$ are shown  for the case with number of frequency channels $N_c= 50$, calibration errors $\Delta_g= 10^{-9}$, instrument noise $\Delta_N=0.01$ $\rm {Jy/{sr}}$, with frequency range from $\nu_{min}=10$ GHz, and $\nu_{max}=1010$ GHz.}\label{Fig:fisher_1}
\end{figure*}

For eleven parameters case, we cannot measure $\mu$ distortion for instrument noise below $2\times 10^{-2}$ Jy/sr (which is about $100$ times better than the capability of the PIXIE design \cite{2014AAS...22343901K}). With an instrument noise of $0.01$ Jy/sr and considering a $10\%$ prior on either $A_{sync}$ or $\alpha_{sync}$, we can make a more than $3$-$\sigma$ measurement of the predicted $\Lambda$CDM $\mu$ distortion signal. The $1$-$\sigma$ marginalized error-bar on the $\mu$ and $y$ parameters are obtained using the Cramer-Rao bound $\sigma_{ii}= (\mathbf{F}^{-1/2})_{ii}$ and are shown in Fig. \ref{fig:SNR_fisher} as a function of the number of frequency channels. For the number of frequency channels below $30$, we are not able to make any measurement of the $\mu$ distortion signal. We need at least $30$ frequency channels to reach a high SNR measurement of the fiducial value of $\mu= 2\times 10^{-8}$ \cite{1994ApJ...430L...5H,Chluba:2016bvg}. The joint $1$-sigma and $2$-$\sigma$  contours for all the eleven parameters are shown in Fig. \ref{Fig:fisher_1}.

A comparison between the IFD technique and  the Fourier transform spectrometer (FTS) can be made by comparing our results with the results from Abitol et al. \cite{Abitbol:2017vwa}. In their analysis \cite{Abitbol:2017vwa}, they have shown the requirements for an FTS in order to distinguish between the spectral distortion signals and foregrounds. They have shown that more than a $3$-$\sigma$ measurement of the fiducial value of the $\mu$ distortion is possible only for instrument noise of  order $0.01$ Jy/sr. From the IFD technique, we are also able to obtain a $6-9$ $\sigma$ measurement of the $\mu$ distortion signal for the number of frequency channels $N_c \geq 50$ with similar instrument noise $0.01$ Jy/sr. This implies that the efficiency of the IFD technique is similar to the standard FTS method in distinguishing between the spectral distortion signal and the foreground contamination. However, the IFD technique provides an additional advantage of combining  the science cases for both imager and spectrometer from a single instrumental setup. This can be a substantial improvement in designing a future mission and exploring the synergies between these two science goals. 
{In reality, the foreground signals are going to require more complex models than the ones we considered here.} The statistical power from a high resolution instrument with the capability to measure the sky spectrum can provide additional benefits {to methods that separate the cosmological and foreground components}. 

\begin{figure*}
\centering
     \includegraphics[trim={1.5cm 1.cm 3.cm 2.cm}, clip, width=1.\textwidth]{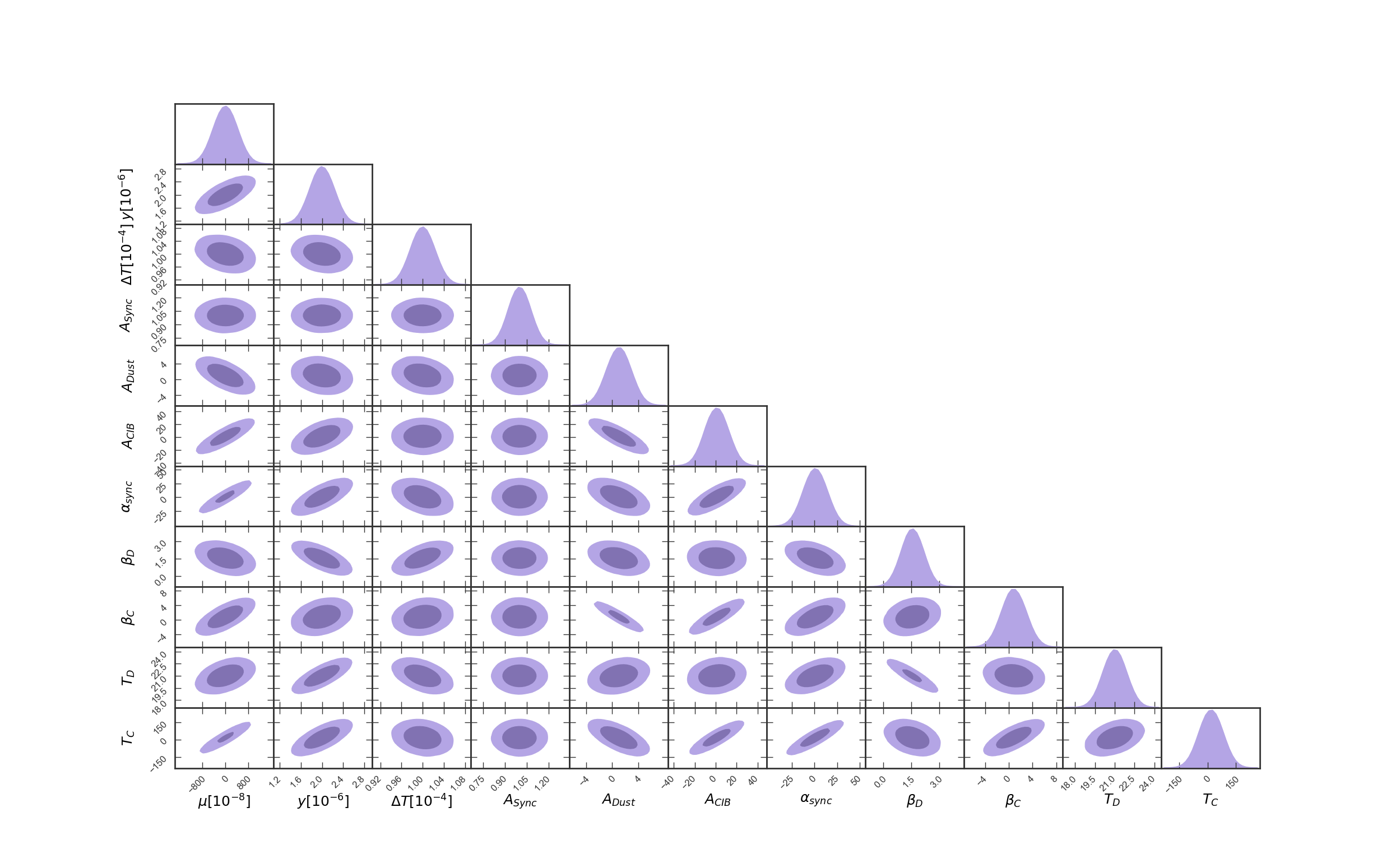}
     \captionsetup{justification=raggedright}
 \caption{Triangle plot: The $1$-$\sigma$ and $2$-$\sigma$ contours  using the Fisher analysis are shown for the case with number of frequency channels $N_c= 40$, calibration errors $\Delta_g= 10^{-6}$, and instrument noise $\Delta_N=10$ $\rm {Jy/{sr}}$, which is similar to a future CMB mission such as PICO \cite{2018SPIE10698E..46Y, 2019arXiv190210541H}. We have used a $30\%$ prior on the $A_{sync}$ parameter for obtaining the results.}\label{Fig:fisher_10jy}
\end{figure*}

\begin{figure*}[t!]
    \centering
    \begin{subfigure}[t]{1.\textwidth}
    \includegraphics[trim={0 0 0 0cm}, clip, width=0.8\textwidth]{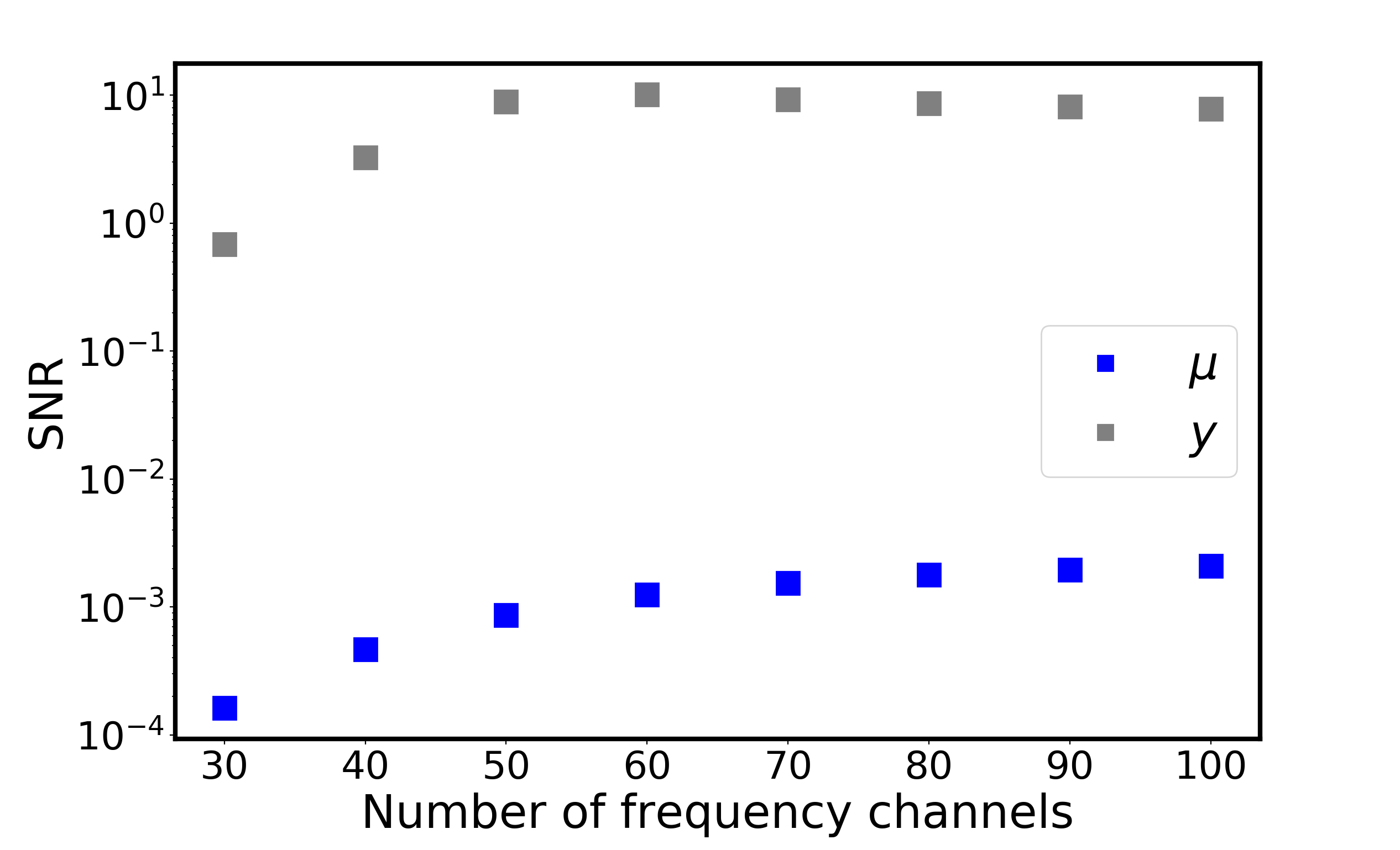}
    \caption{Without prior on any parameters.}
    \end{subfigure}
    \begin{subfigure}[t]{1.\textwidth}
    \includegraphics[trim={0 0 0 0cm}, clip, width=0.8\textwidth]{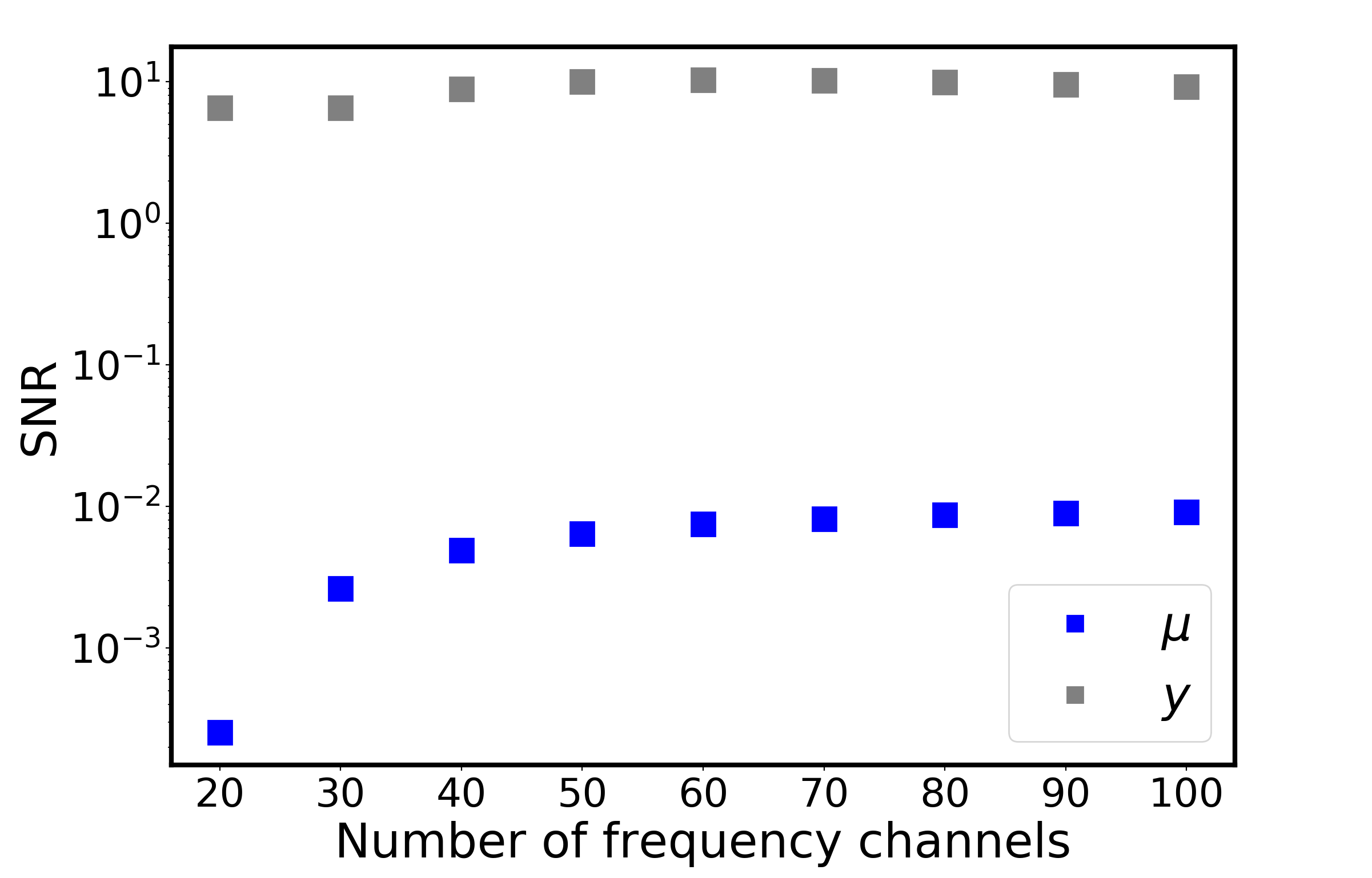}
     \caption{With $30\%$ prior in the amplitude of synchrotron emission $A_{sync}$.}
    \end{subfigure}
    \caption{The signal-to-noise ratio (SNR)($\equiv  (\mu=2\times 10^{-8})/\sigma_{\mu}$ and $(y=2\times 10^{-6})/\sigma_y$)  for $\mu$ and $y$ distortions are shown after marginalizing over all the parameters as a function of the number of frequency channels $N_c$ for the instrument noise  $\Delta_N= 10$ Jy/sr and $\Delta_g= 10^{-6}$. The instrument noise is comparable with the instrument noise of the proposal of a future CMB mission PICO \cite{2018SPIE10698E..46Y, 2019arXiv190210541H}.}
    \label{fig:SNR_fisher_10jy}
\end{figure*}

{We scrutinize the scope of the IFD technique for the instrument noise being  considered for  future CMB missions such as  Probe of Inflation and Cosmic Origins (PICO) \cite{2018SPIE10698E..46Y, 2019arXiv190210541H}. Using an instrument noise of $10$ Jy/sr and an absolute calibrator with $\Delta_g= 10^{-6}$, we have shown the error bars for eleven parameters with number of frequency channels $N_c=\,40$ in Fig. \ref{Fig:fisher_10jy}. We have included a $30\%$  prior on the foreground parameter $A_{sync}$ for this plot. The variation of the SNR  with the number of frequency channels is shown in Fig. \ref{fig:SNR_fisher_10jy}. If a $30\%$ prior can be imposed on the value of the synchrotron amplitude $A_{sync}$, then about a $5$-$\sigma$ measurement of the $y$-distortion is possible  with only $20$ frequency channels between $10-1000$ GHz. If an additional prior on the value of $A_{sync}$ is not possible, then we need about $50$ frequency channels to measure the $y$-distortion with an SNR of about ten. 
Both the plots (Fig. \ref{Fig:fisher_10jy} and Fig. \ref{fig:SNR_fisher_10jy}) indicate that with  PICO-like instrument noise \cite{2018SPIE10698E..46Y, 2019arXiv190210541H}, we can make a measurement of the $y$-distortion at about $10$-$\sigma$. The fiducial value of the $\mu$-distortion ($\mu= 2\times 10^{-8}$) cannot be measured without  orders of magnitude improvement in the calibrator and instrument noise. But using the IFD technique, a PICO-like mission can impose nearly two orders of magnitude stronger constraints than the current bound on $\mu$ from FIRAS \cite{Fixsen:1996nj}. Apart from these advantages , there  are also going to be additional gains of the IFD technique for doing  the science cases of an imager. A more detailed study in the mission concept of PICO \cite{2018SPIE10698E..46Y, 2019arXiv190210541H} to explore the IFD technique would be beneficial for obtaining both imager and spectral distortion science goals.} 

{\section{Challenges in the IFD technique and its comparison with an FTS}}\label{challenges}
{In this section, we compare possible challenges for  the measurement of the spectral distortion signal via the IFD methodology. We compare this with the the FTS method for an experiment such as PIXIE \cite{2011JCAP...07..025K, 2014AAS...22343901K}.} 
\begin{enumerate}
  \item{{Detector sensitivity and instrument noise: The noise of the bolometer detectors \cite{Mather:82} are now limited by the photon noise, with the typical value of the Noise Equivalent Power (NEP) of about few  $\times$ aW$/\sqrt{Hz}$ for the frequency range of interest for the CMB observations. The IFD technique can be implemented with the currently available bolometer detectors proposed for future CMB experiments  \cite{2018SPIE10698E..46Y, 2019arXiv190210541H} with a large number ($\sim10^3$) of detectors. We have shown in Sec. \ref{foregrounds} that with  currently feasible instrument noise (such as proposed for  PICO \cite{2018SPIE10698E..46Y, 2019arXiv190210541H}), we can achieve a high SNR measurement of the $y$ signal. For the measurement of the fiducial value ($2\times 10^{-8}$) of $\mu$-distortion with an SNR of $10$, we need to achieve lower noise, particularly for the frequency channels below $200$ GHz. One of the possible ways to achieve lower instrument noise is to go towards a measurement scheme with multiple numbers of modes $n_m$, rather than single mode measurements. The number of modes depends upon the etendue ($A\Omega$) of the optical system and the wavelength of the photon by the relation $n_m= A\Omega/\lambda^2$. With an increase in the number of modes, we can reduce the noise as $1/\sqrt{n_m}$. For comparison with a mission using FTS, a measurement of the $\mu= 2 \times 10^{-8}$ with an SNR of three also requires an instrument noise $\sim 0.01$ Jy/sr, after adding a $10\%$ prior on the foreground parameters related to the synchrotron emission \cite{Abitbol:2017vwa}. 
  The IFD method is able to achieve an SNR (about $8-10$) with $10\%$ prior on the $A_{sync}$ parameter, even after including the uncertainty from calibration errors (see Fig. \ref{fig:SNR_fisher}). The ability for an FTS instrument design to achieve $0.01$ Jy/sr is yet to be shown and is beyond the reach of the PIXIE mission  \cite{2011JCAP...07..025K, 2014AAS...22343901K}. Recently, the feasibility of reaching this noise level has been discussed \cite{2019arXiv190901593C}.}}
    
    \item{{Absolute calibrator: Measurement of the spectral distortion signals require an absolute blackbody calibrator. The sensitivity of the calibrator decides the measureability of the distortion signal and the ability to remove the foreground contamination. As shown in Sec. \ref{foregrounds} of this paper, an imager with the IFD technique can measure the $y$-distortion signal at $\sim 10$-$\sigma$ with the use of an absolute calibrator having an uncertainty $\Delta_g= 10^{-6}$. We expect a similar accuracy for the absolute calibrator will also be required for an instrument using the FTS technique.\footnote{{A direct comparison of the required accuracy of the absolute blackbody calibrator for the IFD and FTS technique to distinguish the foreground contamination is not possible due to the unavailability of the error analysis in any of the published proposals of the spectral distortion experiment \cite{2011JCAP...07..025K, 2014AAS...22343901K}.}}}} 
    
    \item{ {Number of frequency channels and band width: The number of frequency channels  required for the IFD technique to distinguish between the signals and the foregrounds is typically above $40$, which for equal bandwidth channels leads to a $\Delta \nu \leq 25$ GHz for the frequency channels between $10$ GHz to $1010$ GHz. In the IFD technique, different values of the $\Delta \nu$ can be chosen for different ranges of frequencies. In the case of an FTS instrument, the channel bandwidth is fixed for all the frequency channels by the mirror stroke \cite{2011JCAP...07..025K, 2014AAS...22343901K}, and cannot be chosen separately for different  frequency ranges.}}
    
    \item{{Band pass filters: The band pass filters for each frequency channel with a bandwidth $\Delta \nu$ is required for the IFD technique in order to select operational frequency. For negligible time variation of the filters during the operational period of the mission, each frequency channel can be calibrated from ground before} {the launch of the mission. Along with the pre-mission characterization of the filters, the bands can also be calibrated with the absolute blackbody calibrator during the operation of the mission. In order to implement the IFD technique, it is essential that the same setup of the detectors and the filters see the sky as well as the absolute blackbody calibrator in the same manner throughout the mission. In this way, the modelling of the observed sky signal is going to incorporate the bandpass filter according to Eq. \eqref{skymodel-4}. For an FTS \cite{2011JCAP...07..025K, 2014AAS...22343901K}, the frequency bandpass depends upon the fringe sampling and apodization, which can be determined with the required accuracy.}}
    
   \item{ {Removal of foreground contamination: Foreground contamination is  one of the primary hurdles to overcome in order to detect the spectral distortions in the CMB blackbody. The strength of the foreground contamination is stronger than the fiducial value of the spectral distortion signals {in all of} the frequency channels considered in this analysis. The ability of the IFD technique to distinguish the spectral distortion signal is feasible  as long as the spectral shapes of the foreground contamination and spectral distortion signals are different. It is a challenge for both the FTS and the IFD techniques to successfully remove the foregrounds in order to measure the spectral distortion signal.  { As we have discussed in Sec. \ref{foregrounds}, the IFD technique is capable of distinguishing  spectral distortion signals from foregrounds similar to those with an FTS.}\footnote{{A detailed analysis showing the requirements for a FTS to distinguish between spectral distortion signals and foreground contamination are shown by Abitol et al.\cite{Abitbol:2017vwa}.}} We show in Sec. \ref{foregrounds} that for  PICO-like instrument noise, we can measure $y$-distortion signal with an SNR of ten after marginalizing over all parameters and without using any prior on any of these parameters.}}
\end{enumerate}

\section{\textbf{Conclusions}}
One   can  measure  guaranteed CMB spectral distortions by using \textit{an imaging telescope along with an   inter-frequency calibrator}. This method optimizes  the \textit{inter-frequency differential technique} (difference of the sky intensity between a pair of frequency channels)  to 
explore the shape of the different spectral distortion signals and astrophysical foregrounds. 

We applied the numerical setup InCAS-MC to $600$ instrument configurations  to obtain the joint posterior of six spectral distortion and astrophysical foreground parameters. 
{We  demonstrate that  the expected SNR can be achieved for measuring  hitherto unprecedented  $y$ and $\mu$ distortions from an instrument with PICO-like noise, if the proposed IFD technique can be implemented into the PICO design.  A PICO-like mission could make a $10$-$\sigma$ measurement of the expected $y$-distortion  signal at high angular resolution and  impose constraints on the $\mu$-distortion some  two orders of magnitude below the current bounds from FIRAS \cite{Fixsen:1996nj}.}

{More futuristically, but very much within current planning constraints, for example for the ESA Voyage 2035-2050 program, \footnote{\url{https://www.cosmos.esa.int/web/voyage-2050}} \cite{2019arXiv190901591D, 2019arXiv190901593C}
we show that a CMB imaging telescope with instrument noise $\Delta_N \leq 0.1$ Jy/sr and an   inter-frequency calibrator with calibration error $\Delta_g \leq 10^{-9}$ can make a robust and high SNR detection of the $\mu$, $y$ distortion with a number of frequency channels $20$--$100$.}  For these instrument setups and foreground parameters, we can make measurements of $\mu$ and $y$ distortions with $1$--$\sigma$ error-bars of (a few) $\times 10^{-9}$ as shown in Table. \ref{tab:1}.
For 10 times larger errors in  instrument noise, but with the same $\Delta_g $, measurement of $\mu=2\times 10^{-8}$ will not be possible at more than $~1.5$--$\sigma$. But the $y$ parameter can be measured with high SNR ($\sigma_y= 3.3 \times 10^{-9}$). We  find that for values of $\Delta_g \geq 10^{-9}$, the mean value of the inferred signal will be biased for both $\mu$ and $y$.   
Our findings for different instrument configurations are summarized in Table \ref{tab:1}. {Also for a {more general} foreground model, the IFD technique can distinguish between the foregrounds and spectral distortion signal. With an eleven parameters Fisher analysis, we have shown that the $\mu$ distortion can be distinguished from the synchrotron spectral $\alpha_{sync}$ index and amplitude $A_{sync}$ for instrument noise below $0.02$ Jy/sr. This finding is also similar to the results of the  analysis for an FTS \cite{Abitbol:2017vwa}. 
We emphasize that we are not suggesting here any  technological means for obtaining an inter-frequency  calibration accuracy of $10^{-9}$. This is beyond the scope of this paper, although  we maintain that this is an experimental option that should not be foreclosed at this stage of CMB imaging telescope design.  Nor do we hold any illusions about the simplicity of the dust modelling that we have incorporated into our forecasts. Improved models for foregrounds can surely be incorporated into our scheme, especially with the aid of new data from ongoing surveys such as C-BASS \cite{Jones:2018lvt} which will be beneficial to understand the synchrotron emission. Our point is simply to show that  if future design studies enable implementation of improved calibration and foreground modelling beyond what is currently  envisaged, then
{\it  an imaging telescope with an inter-frequency calibrator is capable of measuring  CMB spectral distortion signals at a level that is up to 5 orders of magnitude more sensitive than FIRAS.} 

\paragraph*{Acknowledgements}
S.M. and B.D.W are supported by the Labex ILP  ANR-10-LABX-63), part of IDEX SUPER, and received financial state aid from the Agence Nationale de la Recherche, as part of the programme Investissements d'Avenir under  ANR-11-IDEX-0004-02. B.D.W. is also supported by the Simons Foundation and the grant ANR-16-CE23-0002. B.D.W thanks the CCPP at New York University for its hospitality while this work was completed. {This work has  used of the Horizon Cluster hosted by Institut d'Astrophysique de Paris. We thank Stephane Rouberol for smoothly running this cluster. We acknowledge the use of Corner \cite{corner}, emcee: MCMC Hammer \cite{2013PASP..125..306F},  Giant-Triangle-Confusogram \cite{Bocquet2016}, IPython \cite{PER-GRA:2007}, Matplotlib \cite{Hunter:2007},  NumPy \cite{2011CSE....13b..22V}, and SciPy \cite{scipy}.

\bibliography{fsd_2_descriptive}
\bibliographystyle{apsrev}
\label{lastpage}
\end{document}